\documentclass[aps,prl,twocolumn,groupedaddress]{revtex4}
\usepackage{amsmath}
\newcommand{\bpsi}{\bar{\psi}}
\newcommand{\bxi}{\bar{\xi}}
\newcommand{\pd}{\partial}
\newcommand{\be}{\stackrel{!}{=}}

\begin{document}

\title{Constructing a supersymmetric generalization of the Gross-Neveu model}

\author{Christian Fitzner}

\affiliation{Institut f\"ur Theoretische Physik III, Universit\"at Erlangen-N\"urnberg}

\date{\today}

\begin{abstract}
A class of 1+1 dimensional supersymmetric theories with four-fermionic interaction will be built from scratch. The vacua of selected examples will be examined in the 't Hooft limit and compared to the Gross-Neveu model.
\end{abstract}


\maketitle

\section{Introduction}
In this paper I want to generalize the Gross-Neveu model to a supersymmetric theory. The questions I try to answer are: is there such a theory? And if yes, how does it compare to the standard Gross-Neveu model?

The Gross-Neveu-Model\cite{GN} is a renormalizable relativistic field theory in 1+1 dimensions that has a four-fermion interaction (``$\psi^4$ theory''), shows asymptotic freedom and is analytically soluble in the \mbox{'t Hooft} limit ($N\to\infty$ while keeping $Ng^2$ constant with $N$ the number of flavours and $g^2$ the coupling constant of the $\psi^4$ interaction) even at finite temperature and density (\cite{Schon:2000qy} provides an overview). 

Supersymmetry (SUSY), the transformation of fermions into bosonic partners ($\delta_\xi\psi\propto\Phi\xi$) and vice versa ($\delta_\xi\Phi\propto\psi\xi$), is in 3+1 dimensions a candidate for physics beyond the standard model. Even (softly) broken it would solve the hierarchy problem in the Higgs sector (why do bare mass and quantum corrections compensate to a physical mass several orders of magnitude smaller?) and provide a possible dark matter particle. While in 3+1 dimensions supersymmetric $\psi^4$ theories have been looked at (e.g. SNJL in \cite{BuLove}), supersymmetric theories in 1+1 dimensions in general concentrate on interactions terms of the form $\bpsi\psi V(\Phi)+V^2(\Phi)$. A thorough introduction into this field is given by \cite{Aitchison} while \cite{Wipf} provides additional information on techniques for low dimensions and the use of Majorana spinors. 

This paper follows my diploma thesis(\cite{Fitzner}). The first is dedicated to the formulation of general supersymmetric theories in 1+1 dimensions. Instead of using the elegant but very formal superfield ansatz I will build the theory from the free theory and check all classes of interaction terms for supersymmetric invariance by hand. In the second part I will select the most Gross-Neveu like theories and examine their vaccuum in the \mbox{'t Hooft} limit.

\section{Building the supersymmetric theory}
\subsection{Basics and free theory}
The theory is formulated in the usual way with Majorana spinors and real scalar fields. For the Dirac matrices the following Majorana representation is used
\begin{flalign*}
\gamma^0&=\sigma_2,\quad\gamma^1=i\sigma_1,\quad\gamma_5=-\gamma_0\gamma_1=\sigma_3.&&
\end{flalign*}
In this representation for Majorana spinors and their bilinears the following relations hold true
\begin{flalign*}
\psi^*&=\psi,&&\\
\bxi\psi&=\bpsi\xi, \quad\bxi\gamma^\mu\psi=-\bpsi\gamma^\mu\xi, \quad\bxi\gamma_5\psi=-\bpsi\gamma_5\xi,&&\\
2\psi\bxi&=-(\bxi\psi)-\gamma_\mu(\bxi\gamma^\mu)-\gamma_5(\bxi\gamma_5\psi)\quad\textrm{(Fierz identity)}.&&
\end{flalign*}

The Lagrangian of the free theorie with a scalar field $\Phi$, a Majorana field $\psi$ and an auxiliary scalar field $F$, that is needed to match bosonic and fermionic off-shell degrees of freedom and can be eliminated via the Euler-Lagrange equation, reads
\begin{flalign*}
\mathcal{L}&=(\pd_\mu\Phi)(\pd^\mu\Phi) +i\bpsi\gamma^\mu\pd_\mu\psi + F^2&&
\end{flalign*}
and is invariant under the following SUSY transformations
\begin{flalign*}
\delta_\xi\Phi&=\bpsi\xi,&&\\
\delta_\xi\psi&=-i(\pd_\mu\Phi)\gamma^\mu\xi-F\xi,&&\\
\delta_\xi F&=-i(\pd_\mu\bpsi)\gamma^\mu\xi,&&
\end{flalign*}
where the parameter $\xi$ is a constant Majorana spinor. These transformations are determined by demanding linearity in parameter and fields, correct Lorentz transformations and correct mass dimensions. Introducing flavours, necessery to obtain $\psi^4$ interactions and labelled by $a=1\ldots N$, yields
\begin{flalign*}
\mathcal{L}&=(\pd_\mu\Phi_a)(\pd^\mu\Phi_a) +i\bpsi_a\gamma^\mu\pd_\mu\psi_a + F_aF_a,&&\\
\delta_\xi\Phi_a&=\bpsi_a\xi,&&\\
\delta_\xi\psi_a&=-i(\pd_\mu\Phi_a)\gamma^\mu\xi-F_a\xi,&&\\
\delta_\xi F_a&=-i(\pd_\mu\bpsi_a)\gamma^\mu\xi.&&
\end{flalign*}
\subsection{Interaction terms}
To obtain a theory with interactions I collect all possible combinations of the fields that are Lorentz scalar and invariant under $O(N)$ flavour transformations and sort them by mass dimension. Then I try to find a set of coefficients that yields a SUSY invariant Lagrangian density. 
\subsubsection{$\mathcal{O}(M)$ interaction terms}
The possible field combinations with mass dimension $M^1$ (and consequently interaction terms with a coupling of dimension $M^1$) are
\begin{flalign*}
(a)&=\bpsi_a\psi_a(\Phi)^{2l_a},&&\\
(b)&=\Phi_a\bpsi_a\psi_b\Phi_b(\Phi)^{2l_b},&&\\
(c)&=\Phi_a F_a(\Phi)^{2l_c}.&&
\end{flalign*}
The factor $(\Phi)^{2l}:=(\Phi_a\Phi_a)^l$ stems from the fact that every term can be multiplied by arbitary powers of $\Phi_a\Phi_a$ without changing mass dimension or behaviour under Lorentz and flavour transformations. Calculating the SUSY transformations and using the Fierz identity to simplify some of the terms, I obtain
\begin{small}
\begin{flalign*}
\delta_\xi(a)=&2i(\pd_\mu\Phi_a)\bxi\gamma^\mu\psi_a(\Phi)^{2l_a} -2F_a\bxi\psi_a(\Phi)^{2l_a} +&&\\ &+2l_a\bpsi_a\psi_a\bxi\psi_b\Phi_b(\Phi)^{2l_a-2},&&\\
\delta_\xi(b)=&2i(\pd_\mu\Phi_a)\Phi_a\Phi_b\bxi\gamma^\mu\psi_b(\Phi)^{2l_b}+2\bpsi_a\psi_b\bxi\psi_a\Phi_b(\Phi)^{2l_b} - &&\\
 &-2F_a\Phi_a\Phi_b\bxi\psi_b(\Phi)^{2l_b} +&&\\
&+ 2l_b\bpsi_a\psi_b\bxi\psi_c\Phi_a\Phi_b\Phi_c(\Phi)^{2l_b-2}=&&\\
=&2i(\pd_\mu\Phi_a)\Phi_a\Phi_b\bxi\gamma^\mu\psi_b(\Phi)^{2l_b} - \bpsi_a\psi_a\bxi\psi_b\Phi_b(\Phi)^{2l_b} -&&\\
&-2F_a\Phi_a\Phi_b\bxi\psi_b(\Phi)^{2l_b},&&\\
\delta_\xi(c)=&i\Phi_a\bxi\gamma^\mu(\pd_\mu\psi_a)(\Phi)^{2l_c} + F_a\bxi\psi_a(\Phi)^{2l_c} +&&\\
&+ 2l_cF_a\Phi_a\Phi_b\bxi\psi_b(\Phi)^{2l_c-2}. &&
\end{flalign*}
\end{small}
Calculating the interaction Lagrangian density for a fixed number of fields (since supersymmetry does not change the total number of fields) and collecting the same combinations of fields and derivatives yields
\begin{small}
\begin{flalign*}
\delta_\xi\mathcal{L}_{1}^{(l)} =& \beta_a(a) + \beta_b(b) + \beta_c(c) =&&\\
=& 2i\beta_a(\pd_\mu\Phi_a)\bxi\gamma^\mu\psi_a(\Phi)^{2l_a}+&&\\
&+i\beta_c\Phi_a\bxi\gamma^\mu(\pd_\mu\psi_a)(\Phi)^{2l_c}  +&&\\
&+ 2i\beta_b(\pd_\mu\Phi_a)\Phi_a\Phi_b\bxi\gamma^\mu\psi_b(\Phi)^{2l_b}+&&\\
&+\Big(-2\beta_a(\Phi)^{2l_a} +\beta_c(\Phi)^{2l_c}\Big)F_a\bxi\psi_a +&&\\
&+ \Big(2l_a\beta_a(\Phi)^{2l_a-2} - \beta_b(\Phi)^{2l_b}\Big)\bpsi_a\psi_a\bxi\psi_b\Phi_b + &&\\
&+\Big(-2\beta_b(\Phi)^{2l_b} +2l_c\beta_c(\Phi)^{2l_c-2}\Big) F_a\Phi_a\Phi_b\bxi\psi_b. &&
\end{flalign*}
\end{small}
The first three terms can be combined to a total derivative for $\beta_a=\frac{1}{2}\beta_c$, $\beta_b=l_c\beta_c$ and $l_a=l_b+1=l_c$. The last three terms vanish with the same relations of $\beta_a,\beta_b,\beta_c$. The complete Lagrangian density of interaction terms of mass dimension $M^1$ can be written as
\begin{flalign*}
\mathcal{L}_{int}^{M}=&\big(\tfrac{1}{2}\bpsi_a\psi_a+F_a\Phi_a\big)W_1 + \Phi_a\bpsi_a\psi_b\Phi_bW'_1&&\\
\textrm{with}&&&\\
W_1(\Phi^2)&:=\sum_{l=0}^{\infty}m_l(\Phi_a\Phi_a)^l,\quad\quad W'_1(\Phi^2):=\frac{\pd W_1}{\pd(\Phi^2)}.&&
\end{flalign*}
Combining the free theory with an interaction Lagrangian given by $W_1=-m$ results, after eliminating the auxiliary field, in the Lagrangian of free massive scalar and Majorana fields with mass $m$.

\subsubsection{$\mathcal{O}(M^2)$ interaction terms}
All possible interaction terms with massless coupling (and field combinations with dimension $M^2$) are
\begin{flalign*}
(1)&=\bpsi_a\psi_a\bpsi_b\psi_b(\Phi)^{2n_1}\quad\textrm{(the interesting term)},&&\\
(2)&=\bpsi_a\psi_b\bpsi_c\psi_c\Phi_a\Phi_b(\Phi)^{2n_2},&&\\
(\bar{1})&=F_aF_a(\Phi)^{2\bar{n}_1},&&\\
(\bar{2})&=F_aF_b\Phi_a\Phi_b(\Phi)^{2\bar{n}_2},&&\\
(\hat{1})&=F_a\Phi_a\bpsi_b\psi_b(\Phi)^{2\hat{n}_1},&&\\
(\hat{2})&=F_a\Phi_b\bpsi_a\psi_b(\Phi)^{2\hat{n}_2},&&\\
(\hat{3})&=F_a\Phi_a\Phi_b\Phi_c\bpsi_b\psi_c(\Phi)^{2\hat{n}_3},&&\\
(A)&=\Phi_a(\pd_\mu\Phi_a)\Phi_b(\pd^\mu\Phi_b)(\Phi)^{2n_A},&&\\
(B)&=\Phi_a(\pd_\mu\Phi_b)\Phi_a(\pd^\mu\Phi_b)(\Phi)^{2n_B},&&\\
(C)&=(i\bpsi_a\gamma^\mu\pd_\mu\psi_a)\Phi_b\Phi_b(\Phi)^{2n_C},&&\\
(D)&=(i\bpsi_a\gamma^\mu\pd_\mu\psi_b)\Phi_a\Phi_b(\Phi)^{2n_D},&&\\
(G)&=i\bpsi_a\gamma^\mu\psi_b(\pd_\mu\Phi_a)\Phi_b(\Phi)^{2n_G}.&&
\end{flalign*}
Using the same procedure as in the $M^1$ case (resulting in 29 equations for the relations of 12 coefficients instead of 4 for 3, full calculation in \cite[ch. 3.4]{Fitzner}) I obtain
\begin{flalign*}
&\mathcal{L}_{int}^{M^2}=&&\\
=&\Big((\pd_\mu\Phi_a)(\pd^\mu\Phi_a) + i\bpsi_a\gamma^\mu(\pd_\mu\psi_a) +F_aF_a\Big)W_2 +&&\\
&+\Phi_a(\pd_\mu\Phi_a)\Phi_b(\pd^\mu\Phi_b)W'_2 + i\bpsi_a\gamma^\mu(\pd_\mu\psi_b)\Phi_a\Phi_bW'_2 +&&\\ &+i\bpsi_a\gamma^\mu\psi_b(\pd_\mu\Phi_a)\Phi_bW'_2+\Phi_aF_aF_b\Phi_bW'_2+&&\\
&+ 2F_a\bpsi_a\psi_b\Phi_bW'_2 - \tfrac{1}{4}\bpsi_a\psi_a\bpsi_b\psi_bW'_2 -&&\\
&-\tfrac{1}{2}\bpsi_a\psi_a\Phi_b\bpsi_b\psi_c\Phi_cW''_2 + F_a\Phi_a\Phi_b\bpsi_b\psi_c\Phi_cW''_2,&&\\
&W_2(\Phi^2):= \sum_{n=0}^{\infty}\lambda_{n}\Phi^{2n},\quad W'_2:=\frac{\pd W_2}{\pd(\Phi^2)}.&&
\end{flalign*}
The case $W_2=\frac{1}{2}$ reproduces the free theory, $W_2=\frac{1}{2}-g^2\Phi_b\Phi_b$ yields the most simple supersymmetric theory with $\psi^4$ interaction.

\section{Examining the theory in the 't Hooft limit}
To obtain the \mbox{'t Hooft} limit I first eliminate the auxiliary field $F$, then take the usual Euler-Lagrange equations for $\psi$ and $\Phi$, replace all flavour singlets by their vacuum expectation values and use the fact that these vanish for field combinations that are not Lorentz scalar. The remaining expectation values are
\begin{flalign*}
\langle\Phi_b\Phi_b\rangle&:=N\sigma_0, &\langle(\pd_\mu\Phi_b)(\pd^\mu)\Phi_b\rangle&:=E_0,&&\\
\langle\bpsi_b\psi_b\rangle&:=N\rho_0, &\langle i\bpsi_b\gamma^\mu\pd_\mu\psi_b\rangle&:=G_0.&&
\end{flalign*}

\subsection{Massive model}
The massive model is given by choosing $W_1=-m_0$ and $W_2=\frac{1}{2}-g^2\Phi_b\Phi_b$. The terms containing $F_a$ can be replaced via the Euler-Lagrange equation by 
\begin{flalign*}
\mathcal{L}^F&=\frac{g^4\bpsi_a\psi_a\bar{A}A}{1-2g^2\Phi_b\Phi_b}-2m_0g^2\bar{A}A\frac{1-2g^2\Phi_a\Phi_a}{1-4g^2\Phi_a\Phi_a}+&&\\
&+\frac{8m_0g^6\bar{A}A(\Phi_a\Phi_a)^2}{(1-2g^2\Phi_b\Phi_b)(1-2g^2\Phi_c\Phi_c)}-\frac{m^2\Phi_a\Phi_a}{2(1-4g^2\Phi_b\Phi_b)},&&\\
\intertext{where}
A&:=\Phi_a\psi_a.&&
\end{flalign*}
Only the last term will contribute in the \mbox{'t Hooft} limit. For $\psi$ and $\Phi$ the Euler-Lagrange equations in the \mbox{'t Hooft} limit read
\begin{flalign*}
0&=(1-2Ng^2\sigma_0)i\gamma^\mu\pd_\mu\psi_a + Ng^2\rho_0\psi_a-m_0\psi_a,&&\\
0&=(1-2Ng^2\sigma_0)(\pd^\mu\pd_\mu\Phi_a) + 2g^2(E_0+G_0)\Phi_a +&&\\
&+\frac{m_0^2}{(1-4Ng^2\sigma_0)^2}\Phi_a.&&\\
\intertext{These are equations for free massive fermions and scalar bosons. The masses are}
M_\psi&=\frac{m_0-Ng^2\rho_0}{1-2Ng^2\sigma_0},&&\\
M_\Phi^2&=\frac{2g^2(E_0+G_0)}{1-2Ng^2\sigma_0}+\frac{m_0^2}{(1-2Ng^2\sigma_0)(1-4Ng^2\sigma_0)^2}.&&\\
\intertext{The expectation values can be calculated easily for the free massive case and are:}
Ng^2\sigma_0&=\frac{Ng^2}{2\pi}\ln\frac{\Lambda}{M_\Phi},&&\\
Ng^2\rho_0&=\frac{-M_\psi Ng^2}{\pi}\ln\frac{\Lambda}{M_\psi},&&\\
g^2E_0&=M_\Phi^2Ng^2\sigma_0,&&\\
g^2G_0&=M_\psi Ng^2\rho_0,&&
\end{flalign*}
where $\Lambda$ is a UV cutoff $\gg M_\Phi,M_\psi$. Solving these conditions in general is difficult because $M_\Phi$ and $M_\psi$ have an additional logarithmic relation (given by $Ng^2\rho_0=-2M\psi Ng^2\sigma_0-\frac{Ng^2M_\psi}{\pi}\ln\frac{M_\Phi}{M_\psi}$), leading to a transcendental equation for $\frac{M_\Phi}{M_\psi}$ that also contains the bare coupling $Ng^2$.\\
The problematic term vanishes in the supersymmetric ansatz $M_\Phi\be M_\psi:=M$ and consequently $\rho_0=-2M\sigma_0$. Using this ansatz in the equation for $M_\psi$ yields
\begin{flalign}
M&=\frac{m_0+2MNg^2\sigma_0}{1-2Ng^2\sigma_0}.\quad\Leftrightarrow m_0=M(1-4Ng^2\sigma_0)\label{eq:gapm}&&\\
\intertext{Substitute $m_0$ in equation for $M_\Phi^2$:}
M^2&=\frac{2M^2Ng^2\sigma_0-4M^2Ng^2\sigma_0}{1-2Ng^2\sigma_0}+\frac{M^2}{1-2Ng^2\sigma_0}=M^2.&&\nonumber
\end{flalign}
The gap equation \ref{eq:gapm} is, save for a factor 2 in the coupling, identical to the one of the massive Gross-Neveu model (compare \cite[ch. 3.1.1]{Schon:2000qy}).

\subsection{Massless model}
The Lagrangian density of the massless model is given by $W_1=0$ and $W_2=\frac{1}{2}-g^2\Phi_b\Phi_b$, the Euler-Lagrange equations are derived similarily to the massive case and read in the \mbox{'t Hooft} limit
\begin{flalign*}
0=&(1-2g^2N\sigma_0)i\gamma^\mu\pd_\mu\psi_a +g^2N\rho_0\psi_a,&&\\
0=&(1-2g^2N\sigma_0)\pd_\mu\pd^\mu\Phi_a+2g^2(E_0+G_0)\Phi_a.&&\\
\intertext{The selfconsistency conditions are}
M_\psi&=-\frac{Ng^2\rho_0}{1-2Ng^2\sigma_0},&&\\
M_\Phi^2&=\frac{2g^2(E_0+G_0)}{1-2Ng^2\sigma_0}=\frac{2g^2(NM_\Phi^2\sigma_0+NM_\psi\rho_0)}{1-2Ng^2\sigma_0}=&&\\
&=\frac{2Ng^2\sigma_0M_\Phi^2}{1-2Ng^2\sigma_0}-2M_\psi^2.&&\\
\intertext{Here the supersymmetric ansatz yields}
0&=(1-4Ng^2\sigma_0)M&&\\
0&=(3-8Ng^2\sigma_0)M^2\quad\quad\Rightarrow M=0.&&
\end{flalign*}
There is no dynamical generation of a physical mass in the case of preserved supersymmetry.\\ 
The only other selfconsistent solution where both masses are independent of the cutoff is $M_\psi=0$, $\frac{2Ng^2}{\pi}\ln\frac{\Lambda}{M_\Phi}=1$. This case provides a gap equation similiar to the Gross-Neveu model but the supersymmetric case is energetically favoured ($\epsilon=\frac{NM_\Phi^2}{16\pi}$ against $\epsilon_{SUSY}=0$). In both cases the scalar density of the fermions is $\rho_0=0$.

\section{Conclusions}
There is a whole class of renormalizable supersymmetric invariant field theories in 1+1 dimensions with a $O(N)$ flavour symmetry that can be considered as generalizations of the Gross-Neveu model. Examining the vacua in the \mbox{'t Hooft} limit for the most simple of theses generalizations yields the following results:\\
In the massive case there is a solution that shows the same behaviour as the respective Gross-Neveu model:  The effective theory is that of $N$ free scalar and Majorana fields with physical mass $M$ where the relation of coupling $Ng^2$, UV cutoff $\Lambda$, bare mass $m_0$ and $M$ is given by the gap equation $1=\frac{m_0}{M}+\frac{2Ng^2}{\pi}\ln\frac{\Lambda}{M}$.\\
In the massless case the behaviour differs from the original Gross-Neveu model: No physical mass is dynamically generated in the supersymmetric case. This reproduces the result of Buchm\"uller and Love\cite{BuLove} for the NJL model in 3+1 dimensions: the supersymmetry protects the (discrete) chiral symmetry of the original Lagrangian density.\\

\appendix

\begin{acknowledgments}
\section{Acknowledgements}
I want to thank Prof. Thies for his support during the work on my diploma thesis.\\
This work has been supported in part by the DFG under grant TH 842/1-1.
\end{acknowledgments}

\bibliography{cf_susy_gn}

\begin{thebibliography}{6}
\expandafter\ifx\csname natexlab\endcsname\relax\def\natexlab#1{#1}\fi
\expandafter\ifx\csname bibnamefont\endcsname\relax
  \def\bibnamefont#1{#1}\fi
\expandafter\ifx\csname bibfnamefont\endcsname\relax
  \def\bibfnamefont#1{#1}\fi
\expandafter\ifx\csname citenamefont\endcsname\relax
  \def\citenamefont#1{#1}\fi
\expandafter\ifx\csname url\endcsname\relax
  \def\url#1{\texttt{#1}}\fi
\expandafter\ifx\csname urlprefix\endcsname\relax\def\urlprefix{URL }\fi
\providecommand{\bibinfo}[2]{#2}
\providecommand{\eprint}[2][]{\url{#2}}

\bibitem[{\citenamefont{Gross and Neveu}(1974)}]{GN}
\bibinfo{author}{\bibfnamefont{D.~J.} \bibnamefont{Gross}} \bibnamefont{and}
  \bibinfo{author}{\bibfnamefont{A.}~\bibnamefont{Neveu}},
  \bibinfo{journal}{Phys. Rev. D} \textbf{\bibinfo{volume}{10}},
  \bibinfo{pages}{3235} (\bibinfo{year}{1974}).

\bibitem[{\citenamefont{Schon and Thies}(2000)}]{Schon:2000qy}
\bibinfo{author}{\bibfnamefont{V.}~\bibnamefont{Schon}} \bibnamefont{and}
  \bibinfo{author}{\bibfnamefont{M.}~\bibnamefont{Thies}}
  (\bibinfo{year}{2000}), \eprint{hep-th/0008175}.

\bibitem[{\citenamefont{{Buchm{\"u}ller} and {Love}}(1982)}]{BuLove}
\bibinfo{author}{\bibfnamefont{W.}~\bibnamefont{{Buchm{\"u}ller}}}
  \bibnamefont{and} \bibinfo{author}{\bibfnamefont{S.~T.}
  \bibnamefont{{Love}}}, \bibinfo{journal}{Nuclear Physics B}
  \textbf{\bibinfo{volume}{204}}, \bibinfo{pages}{213} (\bibinfo{year}{1982}).

\bibitem[{\citenamefont{Aitchison}(2007)}]{Aitchison}
\bibinfo{author}{\bibfnamefont{I.}~\bibnamefont{Aitchison}},
  \emph{\bibinfo{title}{Supersymmetry in Particle Physics: An Elementary
  Introduction}} (\bibinfo{publisher}{Cambridge Univ. Press},
  \bibinfo{address}{Cambridge}, \bibinfo{year}{2007}).

\bibitem[{\citenamefont{Wipf}(2000-2001)}]{Wipf}
\bibinfo{author}{\bibfnamefont{A.}~\bibnamefont{Wipf}},
  \emph{\bibinfo{title}{Introduction to supersymmetrie, lecture notes, old
  version}} (\bibinfo{year}{2000-2001}).

\bibitem[{\citenamefont{Fitzner}(2010)}]{Fitzner}
\bibinfo{author}{\bibfnamefont{C.}~\bibnamefont{Fitzner}},
  \bibinfo{type}{Diploma thesis}, \bibinfo{school}{Universit\"at
  Erlangen-N\"urnberg} (\bibinfo{year}{2010}).

\end{thebibliography}

\end{document}